\documentclass[twocolumn,showpacs,amsmath,amssymb,pra,longbibliography]{revtex4-1}

\usepackage[pdftex]{graphicx} 
\usepackage{dcolumn}  
\usepackage{bm}       
\usepackage[usenames,dvipsnames]{color}
\definecolor{URLCOL}{rgb}{0,0.52,0.83} 
\definecolor{LINKCOL}{rgb}{0.05,0.5,0} 
\definecolor{orange}{rgb}{0.6,0.3,0} 
\definecolor{CITECOL}{rgb}{0.25,0,0.48} 
\usepackage{epstopdf}
\usepackage{array}
\usepackage{dcolumn}

\usepackage[pdftex,bookmarks,breaklinks,bookmarksopen,bookmarksnumbered,colorlinks,linkcolor=LINKCOL,linktocpage,citecolor=CITECOL,urlcolor=URLCOL,pdfpagemode=UseOutline,pdftex]{hyperref}

\usepackage{fancyhdr}
\fancypagestyle{titlepage}
{\rhead{Total pages: \pageref{page:end}}}

\usepackage{booktabs}
\usepackage{natbib}

\definecolor{TITLECOL}{rgb}{0.1,0.2,0.7} 
\definecolor{SECOL}{rgb}{0.1,0.2,0.7} 
\definecolor{CONTENTSCOL}{rgb}{0.1,0.2,0.7} 
\definecolor{SSECOL}{rgb}{0.25,0,0.48} 
\definecolor{SSSECOL}{rgb}{0.2,0.08,0.53} 
\definecolor{FINCOL}{rgb}{0.01,0.3,0.07} 

\def\coloredtitle#1{\title{\textcolor{TITLECOL}{#1}}} 
\def\coloredauthor#1{\author{\textcolor{CITECOL}{#1}}} 

\definecolor{URLCOL}{rgb}{0,0.17,0.43} 
\definecolor{LINKCOL}{rgb}{0.05,0.4,0} 
\definecolor{CITECOL}{rgb}{0.35,0,0.48} 

\def\sss{\scriptscriptstyle\rm}

\def\bea{\begin{eqnarray}}
\def\eea{\end{eqnarray}}
\def\ben{\begin{equation}}

\def\een{\end{equation}}
\def\benu{\begin{enumerate}}
\def\enu{\end{enumerate}}
\def\bei{\begin{itemize}}
\def\eei{\end{itemize}}

\def\n{n}

\def\sss{\scriptscriptstyle\rm}

\def\g{_\gamma}

\def\F{_{\sss F}}
\def\L{_{\sss L}}

\def\TF{^{\rm TF}}

\def\bei{\begin{itemize}}
\def\eei{\end{itemize}}
\def\beit{\begin{itemize}}
\def\eit{\end{itemize}}
\def\benu{\begin{enumerate}}
\def\enu{\end{enumerate}}

\def\half{\frac{1}{2}}

\def\F{_{\sss F}}

\def\n{n}

\def\sec#1{\section{\textcolor{SECOL}{#1}}}
\def\ssec#1{\subsection{\textcolor{SSECOL}{#1}}}
\def\sssec#1{\subsubsection{\textcolor{SSSECOL}{#1}}}

\def\Kta{K_2}
\def\Kb{K_1}
\def\Kc{K_0}

\begin{document}

\coloredtitle{Leading corrections to local approximations II (with turning points)}
\coloredauthor{Raphael F. Ribeiro}
\affiliation{Department of Chemistry and Biochemistry, University of California, San Diego, CA, 92093}
\coloredauthor{Kieron Burke}
\affiliation{Department of Chemistry, University of California, Irvine, CA 92697}

\begin{abstract}
Quantum corrections to Thomas-Fermi (TF) theory are investigated for
noninteracting one-dimensional fermions with known uniform semiclassical
approximations to the density and kinetic energy.  Their structure is
analyzed, and contributions from distinct phase space regions (classically-allowed
versus forbidden at the Fermi energy) are derived analytically. 
Universal formulas are derived for both particle numbers and energy components in each region.
For example, in the semiclassical limit, 
exactly $(6\pi{\sqrt 3})^{-1}$ of a particle leaks
into the evanescent region beyond a turning point.
The correct normalization of semiclassical densities is
proven analytically in the semiclassical limit.
Energies and densities are tested numerically
in a variety of one-dimensional
potentials, especially in the limit where TF theory becomes exact.
The subtle relation between the pointwise accuracy of the semiclassical
approximation and integrated expectation values is explored.
The limitations of the semiclassical formulas are also investigated when
the potential varies too rapidly.  The approximations are shown to work for multiple
wells, except right at the mid-phase point of the evanescent regions.
The implications for density functional approximations are discussed.

\end{abstract}

\maketitle


\sec{Introduction}
\label{intro}

While the popularity of density functional theory (DFT) has never been higher \cite{PGB15},
the lack of a systematic approach to the construction of approximate 
exchange-correlation functionals or even
orbital-free kinetic energy functionals remains an outstanding issue
confronted by practitioners and developers of the theory alike.
The closest to a systematic approach might be the decades-long
work of Perdew and co-workers, which recently yielded a highly
promising meta-generalized gradient approximation called SCAN,
but only after 20 years of research, and including norms which
are used to fix parameters in the approximation \cite{SRP15}.

Semiclassical approximations have inspired the development of density
functional methods from the start.  The first density functional approximation
is given by Thomas-Fermi (TF) theory \cite{T27,F28}. 
It may be regarded as a classical limit of quantum mechanics. 
As such, it has been proved to be a universal limit for the
quantum mechanics of nonrelativistic matter \cite{LS73}. 
More recently, it has been conjectured that the analogous statement
in Kohn-Sham DFT, that the local density approximation for both
exchange and correlation, also becomes relatively exact in this limit\cite{BCGP16}.
Therefore, it is unsurprising the most successful density
functional approximations reduce to local density approximations in the limit
where the predictions according to the latter become exact.

Over the past decade \cite{ELCB08, LCPB09,EB09, CLEB10, CSPB11, CGB13, RLCE15, RB15, BCGP16}, our group has pursued the connection between
semiclassical approximations and DFT.  Much of the work can be classified as being
in one of two camps: they are either limited to one-dimension, or have a general scope.
The advantage of one dimension is that semiclassical approximations to
wave functions are long known \cite{W26, K26, B26, La37}. Thus more explicit progress, including
analytic results are possible in 1D, and suggest both the greatest power and 
limitations of this approach more generally.  Earlier work made an even greater
simplification, by employing box boundary conditions to avoid the singularities
associated with turning points \cite{CLEB10}.  More recently, at least in the case of
densities and kinetic energy densities, a semiclassical approximation
was derived \cite{RLCE15} which is uniformly asymptotic in space, i.e., suffers no
singularities, while capturing the leading corrections to TF results at
every point.  A brief account appeared in Ref. \cite{RLCE15}, while a more
detailed mathematical derivation is under review \cite{RB15}. In the current work, we test the recently-derived approximations numerically in a variety of 
situations of relevance to atomic and
molecular systems.  We show that, even when TF theory is surprisingly
accurate for quantities integrated over the entire system (such as the
total energy or its components), the uniform semiclassical approximations
capture the leading-corrections within a given region of space.
We use the pointwise formulas to derive analytic corrections
to the TF energies, and confirm these numerically on a class
of potentials. But we also find that many such contributions 
cancel {\em exactly} between classically-allowed and forbidden regions,
which leads to high accuracy of TF theory for integrated quantities,
despite poor pointwise behavior.

\begin{figure} [htb]
\includegraphics[width=\columnwidth]{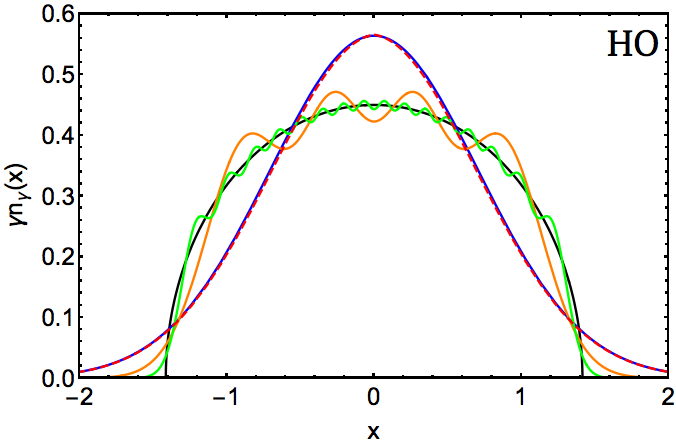}
\caption{$\gamma\, n_{\gamma}(x)$ for $N=1$ harmonic potential where
$\gamma=1$ (blue, with semiclassical approximation dashed red),
1/4 (orange), 1/16 (green), and TF (black).}
\label{ng}
 \end{figure}
To illustrate the main ideas of this paper, in Fig. \ref{ng} we plot
a sequence of densities for same-spin fermions in a harmonic well, $v(x)=x^2/2$. 
In each curve, we replace $\hbar$ by $\gamma\hbar$, $N$ by $N/\gamma$,
and $\n(x)$ by $\gamma\n(x)$,
where $\gamma$ is made successively smaller.  In the limit $\gamma\to 0$,
the exact quantum curve weakly approaches the TF density.
The uniform semiclassical approximation is so accurate as to be
indistinguishable from the exact curves here, even for $N=1$, although
it only includes the leading corrections to TF as $\gamma \rightarrow 0$. 
By any {\em pointwise} measure, it is vastly superior to TF.
Results within this paper demonstrate this for several different potentials.

But DFT cares almost solely about energies \cite{WNJK16}.  To connect the pointwise
success of the uniform approximation with these, we calculate the integrated
densities and energy-densities in forbidden and allowed regions separately.
The semiclassical approximations allow us to derive 
leading corrections to TF in each region analytically,
and check the results numerically.  These are universal formulas that
apply to all (non-pathological) 1D potentials.  Again, the uniform
approximations are vastly superior to TF theory.  However, we also 
show that the improvements in allowed and forbidden regions are always
equal and opposite, and so {\em cancel} from the total energy components.
The harmonic oscillator is a stark example:  because TF
theory yields the exact energy components for this case, the semiclassical approximation always {\em worsen} those energies!

This paper is devoted to demonstrating these facts and discussing their
consequences.
We review in Section \ref{gensc} the semiclassical limit of nonrelativistic fermionic
systems. In Sec.\ref{cast} we establish the adopted notation while providing
a brief discussion of the uniform semiclassical approximations derived 
in Refs. \cite{RLCE15, RB15}. Sec. \ref{method} is devoted to describing both numerical and
analytic calculations.
Section \ref{lead} provides a detailed description of the
leading corrections to TF components in different regions of configuration
space, ranging from pointwise (Sec. \ref{point}) to regional (Sec. \ref{region}) and
finally global (Sec. \ref{glob}). 
In Sec. \ref{beyond}, we study situations that differ qualitatively from the generic
wells studied up to that point.  We see the breakdown of the semiclassical approximation
when the potential varies too rapidly (Sec. \ref{breaksc}), how extended systems
can be treated (Sec. \ref{extend}) and tunneling in a double well(Sec. \ref{tunnel}).
We close with a discussion of the significance of these results, especially in the
context of density functional theory.  The appendix collects useful results about the
Airy functions used in this paper.

\sec{Background}
\label{back}
\ssec{General semiclassical limit of nonrelativistic fermionic systems}
\label{gensc}

\par Lieb and Simon proved in 1973 \cite{LS73} that quantum-mechanical nonrelativistic fermionic systems interacting via the Coulomb interaction are exactly described by TF theory in the semiclassical limit. In particular, this implies the relative error of expectation values predicted by TF theory goes to zero as the nuclear charges $Z$ in the system go to infinity. The reason $Z$ gets involved here is that it sets the relevant length scales for the Coulombic problem \cite{L81}.

More recently, Fournais et al. \cite{FLS15} proved that a generalization of
the Lieb-Simon result is valid in any number of spatial dimensions, i.e.,
 under semiclassical scaling, all correlation functions of a quantum-mechanical system
(and therefore, all of its properties) agree with those obtained by minimization
of the TF energy functional in a well-defined limit. Specifically,
the predictions of TF theory emerge from quantum mechanics when
the number of particles $N$ is scaled to infinity and $\hbar \rightarrow 0$ as 
$N^{-1/d}$, where $d$ is the dimensionality of the considered configuration space.
Hereafter, we restrict considerations to 1D.

\ssec{Relevant classical variables and Thomas-Fermi theory}
\label{cast}

Consider a 1D Hamiltonian
\ben
\hat{h} = -\frac{1}{2}\frac{d^2}{dx^2} + v(x). 
\label{hdef}
\een
We consider potentials that either vanish or diverge positively at large $|x|$.
In the former case, we require at least one bound state for the employed methods to be relevant.
We are interested in the ground-state of $N$ noninteracting same-spin fermions at 0K 
in this potential.   We list the eigenvalues in increasing order, $\epsilon_j$, $j=1,2,...,N.$ Then we may write the particle density as
\ben
n(x)=\sum_{j=1}^{N} |\phi_j(x)|^2, ~~\int_{-\infty}^{\infty}\mathrm{d}x\, \n(x)=N.
\label{ndef}
\een
The kinetic energy may be written many ways, but 
we chose a specific kinetic-energy density, 
\bea
t(x)&=&\sum_{j=1}^{N} \left[\epsilon_j - v(x)\right] |\phi_j(x)|^2,
~~~\int_{-\infty}^{\infty} \mathrm{d}x\, t(x) = T.\nonumber\\
&=&-\half \sum_{j=1}^{N} \phi^*_j(x) \frac{d^2 \phi_j(x)}{dx^2}.
\label{tdef}
\eea
Unlike the particle density, the choice of kinetic energy density is
arbitrary, as the kinetic energy density is not a physical observable.

Our focus is semiclassical approximations, which require classical inputs.
For a given energy $\epsilon$, we consider only the case of two turning points 
(denoted $x_L(\epsilon)$ and $x_R(\epsilon)$)
at which the classical momentum
\ben
k(\epsilon,x) = \sqrt{2[\epsilon - v(x)]},
\label{kdef}
\een
vanishes.  We define the corresponding classical
phase and time 
for the particle to arrive at $x$,
starting from $x_L(\epsilon)$, as:
\bea
\theta(\epsilon,x)&=&\int_{x_L(\epsilon)}^x \mathrm{d}x'\, k(\epsilon,x'),\nonumber\\
\tau(\epsilon,x) &= &
\int_{x_L(\epsilon)}^x  \frac{\mathrm{d}x'}{k(\epsilon,x')}. 
\label{thetadef}
\eea
The classical action for energy $\epsilon$ is determined by the total phase
from left to right turning points:
\ben
I(\epsilon) =  \frac{\theta_L[\epsilon,x_R(\epsilon)]}{\pi}.
\label{Idef}
\een
Inversion yields the energy as a function of the action, $\epsilon(I)$, and
the frequency corresponding to the motion is
\ben 
\omega(\epsilon) = \frac{\pi}{T(\epsilon)} = \frac{d\epsilon}{dI},
\label{omegadef}
\een 
where $ T(\epsilon) = \tau[\epsilon,x_R(\epsilon)]$.
The corresponding angle variable is then 
\ben
\alpha(\epsilon,x)= \omega(\epsilon) \tau(\epsilon,x).
\label{alphadef}
\een
All of the classical observables given above can be obtained
from the classical phase $\theta(\epsilon,x)$ by application of
partial derivatives. For instance,
\bea 
k(\epsilon,x)&=&\frac{\partial \theta(\epsilon,x)}{\partial x}, ~~~
\tau(\epsilon,x)= \frac{\partial \theta(\epsilon,x)}{\partial \epsilon}, \nonumber\\
\alpha(\epsilon,x)&=&\frac{\partial \theta(\epsilon,x)}{\partial I}, ~~~
T(\epsilon)=  \frac{\partial \theta(\epsilon,x_R)}{\partial \epsilon}. 
\eea
The WKB quantization condition for the energy of the $j$-th level is
\ben
I\left(\epsilon_j^\text{WKB}\right)= \hbar \left(j-\half\right).
\label{WKB}
\een
We define the semiclassical Fermi energy for $N$ particles by
\ben
\epsilon_F = \epsilon_{N+1/2}^{\rm WKB} \label{fen},
\een
i.e., we set the quantum number $j$ midway between the index of the highest occupied level
and lowest unoccupied one, 
and use a subscript $F$ to denote quantities evaluated at this energy.  We can also
define the mid-phase point $x_m$ at which 
\ben
\theta\F(x_m) = N \pi/2.
\label{xmdef}
\een
At the Fermi energy, we adopt the convention of always measuring classical variables
from their nearest turning point as measured by the classical phase.  
This leads to small kinks in $\theta_F(x)$ and $\alpha_F(x)$ as $x$ goes
through $x_m$ but which become irrelevant in the semiclassical limit
(see Eq. 36 of Ref. \cite{RB15}). 
For potentials that are symmetric about $x=0$, it follows that $x_m=0$. 
Since one can always treat a right-hand turning point by applying left-hand formulas to
$v(-x)$, we here give only formulas for $x < x_m$.

In 1D it is particularly easy to explicitly connect TF theory
with semiclassical approximations. In the limit where $\hbar\to 0$, the WKB energy for a system
with $N$ occupied orbitals is simply the sum of all occupied semiclassical orbital energies.
On the other hand, by approximating the discrete sum of WKB orbital densities as an integral and ignoring quantum oscillations and exponentially small terms, then it follows that 
\ben
\n(x) = \frac{k\F(x)}{\pi},~~~t(x)= \frac{k\F^3(x)}{6\pi}~~~\text{(PFT)},
\label{TFPFT}
\een
where PFT denotes potential functional theory, i.e., both quantities are given
as functionals of the potential.
Elimination of $k\F(x)$ yields
the TF expression for the 1d kinetic energy density functional:
\ben
t(\n) = \frac{\pi^2 n^3}{6}.  ~\text{(DFT)} \label{tTF}
\een

To study the semiclassical limit of noninteracting 1d fermions \cite{FLS15} we define positive real-valued
parameters $\gamma$, scaled particle number $N_\gamma$, and scaled Planck's constant $h_\gamma$, where, 
\ben
N_\gamma = \frac{N}{\gamma},  ~~~\hbar_\gamma = \gamma\, \hbar. 
\label{gammasc}
\een 
The semiclassical limit arises when $\gamma \rightarrow 0$. This has been
extensively discussed in Refs \cite{ELCB08, CLEB10, RB15, FLS15}. 
For instance, in Ref. \cite{CLEB10}, Cangi et al. derived semiclassical approximations
for the particle and kinetic energy densities of systems with closed (box) boundary
conditions with the requirement that they provide the leading corrections
to TF theory when $\gamma$ is sufficiently small. Recently,
we generalized the approach of Ref. \cite{CLEB10} to the case of 
unbounded domains \cite{RLCE15, RB15}. In Ref. \cite{RLCE15} we provided a
sketch of the construction of  the semiclassical approximations, 
whereas Ref. \cite{RB15} explored the mathematical intricacies of such construction.

Under $\gamma-$scaling the TF densities change trivially:
\ben
\n\TF\g(x) = \frac{k\F(x)}{\pi\gamma},~~~t\TF\g(x)= \frac{k\F^3(x)}{6\pi \gamma},
\label{TFg}
\een
Thus both energy components (kinetic and potential) scale as $\gamma^{-1}$.

While it is easy to find the dominant behavior of any observable in the limit where $\gamma=0$,
i.e., via TF theory, it took about 50 years to find a general
universally valid result for the leading corrections \cite{RLCE15, RB15}.  
The behavior and accuracy of these expressions are relatively unexplored.  We also note that
there is no general procedure for finding the leading corrections to the TF density
functionals of Eq. (\ref{tTF}).  The semiclassical expressions are all functionals of the potential \cite{CLEB11,CGB13, RLCE15}

Going beyond the dominant contribution also means including evanescent behavior.
All of the spatially-varying quantities presented above have a
clear classical meanings for $ x_L(\epsilon) \leq x $. 
However, they become purely imaginary 
in a region which is classically-forbidden. We define
\ben
 -i k(\epsilon,x) = \sqrt{2[v(x)-\epsilon]}, ~x < x_L(\epsilon), 
\label{kevan}
\een
which ensures that $-i \theta$, $-i \tau$ and $-i \alpha$ are real for all $x<x_L$.
With these definitions, all of the previously identified classical quantities are extended
to any $x \in x<x_L$.

\ssec{Uniform semiclassical approximations}
\label{unisemi}
\def\sc{^{sc}}
The Langer wave function is a uniform semiclassical generalization of the WKB
results that remains finite as one passes through a turning point, and is
expressed in terms of Airy functions\cite{La37}. 
By taking the semiclassical limit of finite sums of the squares of
such functions, the main results of Ref. \cite{RLCE15} are obtained\cite{RB15}. 
We write these here in a simple form
\bea
\n\sc(x)&=&f\sc_1[k\F(x),d\F(x),\theta\F(x)],\nonumber\\
t\sc(x)&=&\frac{k\F^2}{6}f\sc_3[k\F(x),d\F(x),\theta\F(x)],
\label{nsc}
\eea
where 
\ben
d\F(x) = k\F(x)\sin\alpha\F(x)/\omega\F,
\label{ddef}
\een
and 
\ben
f\sc_p(k,d,\theta)=
\frac{1}{\pi}\left( k\, \Kc(\theta) +p\frac{\Kb(\theta)}{d}\right),
\label{fsc}
\een
with $K_j$ being known combinations of products of Airy functions 
given in Appendix A, and $d\F(x)$ is a classical measure of distance.
\begin{figure}[htp] 
\includegraphics[width=\columnwidth]{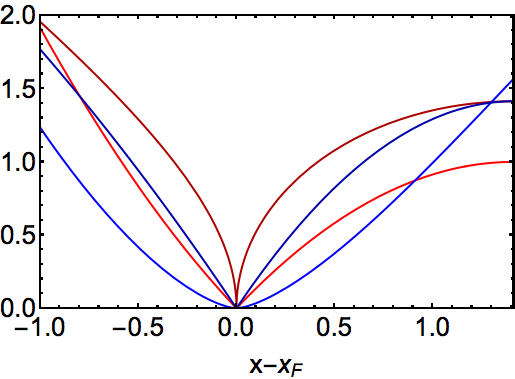}
\caption{Classical properties of $\epsilon_F$ trajectory
for the SHO with $N=1$ vs. $x-x\F$. Red is $\epsilon\F - v(x)$,
dark red is $k_F(x)$, blue is $\theta_F(x)$, and dark blue is $d_F(x)$.}
\label{fcl}
\end{figure}
The input classical variables are plotted in Fig. \ref{fcl}  for a simple
harmonic oscillator, with the origin at the Fermi turning point,
and using the above convention for the evanescent region.  We
see that these  quantities are all comparable in magnitude and
approach the turning point in diverse ways.

The mathematical underpinnings of the above have been discussed in detail in Ref. \cite{RB15}. 
Here we just recall that the analytical continuations given before for the 
spatially-varying classical quantities ensure that the particle density is
continuous and real everywhere in configuration space. It also positive everywhere except
in pathological situations where $v'(x_F) \rightarrow 0$, where $x_F$ is the Fermi turning point.
While the above supposes that $v(x)$ consists of a single potential well, 
we also discuss the generalization to the case of weakly-coupled multiple
wells in Sec \ref{tunnel} below.

In the derivation of Eq. (\ref{fsc}), the first part comes from the
0th-order term in the Poisson summation formula, while the
second is the dominant correction from the oscillating pieces \cite{RLCE15, RB15}.  If $\Kb$ is set to
zero a first approximation to the density is obtained, though it is in general uncontrollable.

\begin{figure}[htp] 
\includegraphics[width=\columnwidth]{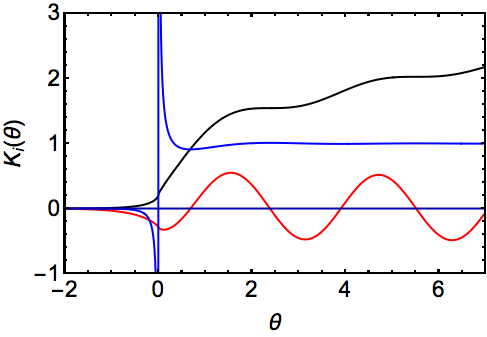}
\caption{$K_i$ functions versus $\theta$, with $K_i$ versus $-i\theta$ in evanescent
region. $\Kc$ is blue, $\Kb$ is red, and $\Kta$ is black.}
\label{Ktheta}
\end{figure}

In Fig. \ref{Ktheta}, we plot the $K_j$ functions against $\theta$, adopting the
convention mentioned above for the evanescent region.   They contain both
the quantum oscillations of the travelling region ($\theta >> \pi$), the singular
behavior near the turning point ($\theta \sim 0$), and the evanescent behavior far from the turning
point ($-i\theta <<0$).  
In the region of the turning point, the different power-law behaviors of the
classical variables balance the singular to produce continuous and
largely smooth behavior in Eq. (\ref{fsc}).
On a large scale, $\Kc \approx H(\theta)$ and $\Kb=0$, where $H$ is the Heaviside
step function.  Inserted in Eq. \ref{nsc}, this yields the TF results.
The function $\Kta$ is useful in Sec \ref{region}
where we explore the small $\gamma$ limit.
\begin{figure} [htb]
\includegraphics[width=\columnwidth]{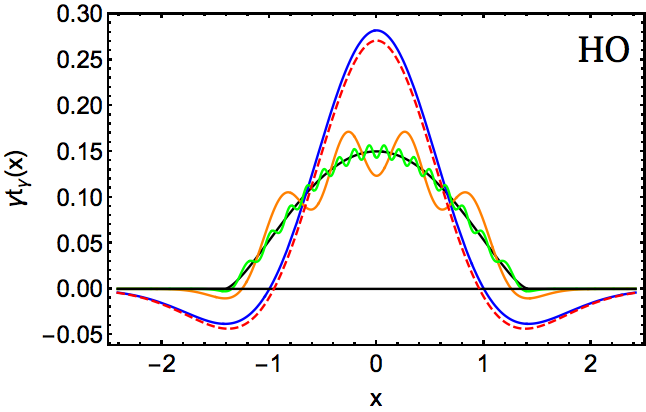}
\caption{Same as Fig. \ref{ng}, but for the kinetic energy density.}
\label{tg}
\end{figure}

The value of this representation becomes clear when we reintroduce $\gamma$.
Then 
\ben
\left\{\begin{array}{l}
\n\sc\g(x)\\
 t\sc\g(x)
\end{array} \right\}
=\frac{f\sc_p(k\F(x),d\F(x),\theta\F(x)/\gamma)}{\gamma},
\label{ntsc}
\een
i.e., beyond the trivial scaling with $\gamma$, only the phase argument depends
on $\gamma$, as all other quantities are purely classical.
As $\gamma\to 0$, only the arguments of the $K_i$ 
functions change, becoming much larger for any fixed $x$.  
Alternatively, the region of $|\theta| < \xi$, for fixed $\xi$,
which we call the turning-point region, shrinks to a region in $x$ space of size $\gamma$.  

In Fig. 1 we only show the result given by the semiclassical formula for $\gamma=1$. For all other values of 
$\gamma$, the semiclassical formula is everywhere indistinguishable from the exact curve.
In Fig. \ref{tg}, we plot the analogous curve for the scaled kinetic energy density, for which the semiclassical approximation is (slightly) less accurate.

\begin{figure}[h]
\includegraphics[width=\columnwidth]{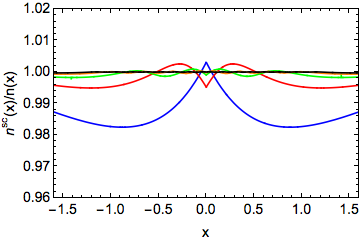}
\caption{$n^{\text{sc}}(x)/n(x)$ for the harmonic oscillator with
$\gamma=1$ (blue), 1/2 (red), 1/4 (green),
1/8 (orange), and 1/16 (black).}
\label{ngratio}
\end{figure}

In Fig. \ref{ngratio}, we plot the ratio of semiclassical and exact densities
for different values of $\gamma$. It appears to  approach 1 everywhere, showing that its relative
error vanishes for sufficiently small $\gamma$, for all values of $x$.  This shows that
it is a {\em uniform} asymptotic expansion and suffers none of the difficulties of patching
for different regions, despite the qualitatively different behavior of
spatially-varying properties in the traveling, transition and
evanescent regions.
In fact, for $|x|$ sufficiently large, the semiclassical density decay is exponential in
$|\theta\F(x)|$ (see Eq. (22) of Ref. \cite{RLCE15}), which does not match the decay of
the exact density,
which is dominated by the highest occupied level, $|\phi_N(x)|^2$.  
Thus the fractional error eventually grows again beyond
the edges of the figure.
But as $\gamma\to 0$, the point at which this error becomes
noticeable becomes ever larger.

These formulas are remarkable for their ability to yield extremely accurate
results using {\em only} classical inputs.  Fig. 1 shows that, for the ground-state
of the harmonic oscillator, they yield densities that are indistinguishable from
the exact quantum curves.  Yet no differential equation has been solved to evaluate
them, and they apply to all potentials.  The smoother the potential is, the more
accurate the result will be, even with only one occupied level.

\sec{Methods}
\label{method}
\ssec{Numerical}
\label{num}
One of the great features of the uniform semiclassical approximations for the particle and kinetic energy densities is that they allow us to obtain these quantities with minimal effort. This is to the contrary of numerically solving the Schrodinger equation for systems with a large number of particles. In this section we explain the numerical methods employed in the paper to compare semiclassical and Thomas-Fermi theory with exact results.

Accurate numerical solutions for the Schrodinger equation were extracted with the Matrix Numerov method \cite{PGW12} whenever the studied potential could not be solved analytically. A grid spacing of $O\left(10^{-3}\right)$ was chosen and the size $L$ of the studied region depended on each potential.
This choice of parameters was guided by the requirement
that both kinetic and total energies converge (relative to both grid spacing and $L$)
up to at least the 3rd decimal digit (except for the double wells where we only enforced convergence up to the 2nd decimal digit). 
In every case we checked that both the exact particle and
kinetic energy densities were at least of $O(10^{-5})$ when $x=\pm L$.
For calculations with $\gamma << 1$, smaller grid spacings were needed to capture the oscillations and the more rapid decay in the evanescent region.
Mathematica 10.1 was employed in all computations \cite{math}.

\ssec{Analytical}
\label{anal}

\def\one{{^{(1)}}}

\begin{table*}[t]
\begin{center}
\begin{tabular}{| l | c | c | c | c |}
\hline
 & harmonic & quartic & Poschl-Teller & Morse\\
\hline
$v(x)$ & $x^2/2$ & $x^4$  & $-D/\text{cosh}^2(ax)$ & $D[\left(1-e^{-a x}\right)^2-1]$\\ 
\hline
$\epsilon_j$ & $\tilde{j}$ & $-$ & $-a^2(\alpha_e-\tilde{j})^2/2$ & $-a^2(\alpha-\tilde{j})^2/2 $\\
$\epsilon_j^{sc}$ & $\epsilon_j$ & $(d_4\tilde{j})^{4/3}$ & $-a^2(\alpha-\tilde{j})^2/2$ &$\epsilon_j$ \\
\hline
$t_j$ & $ \tilde{j}/2$ &  $-$ & $a^2(\alpha^2/\alpha_e - \alpha_e + \tilde{j})(\alpha_e-\tilde{j})/2$ & $a^2\tilde{j}(\alpha-\tilde{j})/2$ \\ 
$t_j^{sc}$ & $t_j$ & $2(d_4\tilde{j})^{4/3}/3$ &$a^2\tilde{j}(\alpha-\tilde{j})/2$  & $t_j$ \\ \hline
$v_j$ & $ \tilde{j}/2$ &  $-$ & $-D(1-\tilde{j}/\alpha_e)$ & $-D(1-\tilde{j}/\alpha)$ \\ 
$v_j^{sc}$ & $v_j$ & $t_j^{sc}/2$ & $-D(1-\tilde{j}/\alpha)$ & $v_j$ \\ \hline
\multicolumn{5}{c|}{Classical energy variables}\\
\hline
$\omega(\epsilon)$ & 1 & $4d_4^{1/3}\epsilon^{1/4}/3$ & $a\sqrt{-2\epsilon}$ &$a\sqrt{-2\epsilon}$\\
$I(\epsilon)$ & $\epsilon$ &  $\epsilon^{3/4}/d_4 $  & $\alpha-\sqrt{-2\epsilon}/a$ & $\alpha-\sqrt{-2\epsilon}/a$  \\
\hline
\multicolumn{5}{c|}{Fermi level}\\
\hline
$\epsilon_F $ & $N$ & $(d_4 N)^{4/3}$ & $-(\alpha a\beta)^2/2$ & $-(\alpha a\beta)^2/2$ \\
$\omega_F$ & $1$ & $4d_4^{4/3} N^{1/3}/3$ &  $a^2 \alpha \beta$ & $a^2 \alpha\beta$ \\
\hline
$x_{F}$ & $-\sqrt{2N}$ & $-d_4^{1/3} N^{1/3}$ & $-\text{cosh}^{-1}(1/\beta)/a$ &$-\text{log}(1 \pm \overline{\beta})/a$ \\
$l_F$ & $2^{-1/2} N^{-1/6}$ & $(d_4 N)^{-1/3}/2$ & $(2\alpha^2 \beta^2 \overline{\beta})^{-1/3}/a$  & $(2\alpha^2 \overline{\beta})^{-1/3}(1\pm \overline{\beta})^{-1/3}/a$\\
$v_F''$ & 1 & $12(d_4 N)^{2/3}$& $\alpha^2 a^4 \beta^2(\beta^2 - 2\overline{\beta}^2)$ & $\alpha^2 a^4(3\pm 3 \overline{\beta}-2\beta^2)$\\
\hline
\multicolumn{5}{c|}{Dimensionless parameters}\\
\hline
$\phi_F$ & $ N^{-2/3}/2$ & $d_4^{2/3} N^{-1/3}/3$ & $(4\alpha \beta \overline{\beta}^2)^{-1/3}$ & $(4\alpha\overline{\beta}^2)^{-1/3}(1\pm\overline{\beta})^{-2/3} \beta$ \\
$\chi_F$ &$N^{-2/3}/4$ & $3(d_4N)^{-2/3}/4$ & $(16 \alpha^2 \beta^2 \overline{\beta}^4)^{-1/3} (\beta^2 - 2\overline{\beta}^2)$ & $ \alpha^{-2/3}[2(\overline{\beta}\pm \overline{\beta}^2)]^{-4/3}(3-2\beta^2\pm 3\overline{\beta})$ \\
$l_F^2 \epsilon_F$ & $N^{2/3}/2$ & $ (d_4 N)^{2/3}/4$ & $-(\alpha^2 \beta^2 \overline{\beta}^{-2}/4)^{1/3}/2$ & $-(\overline{\beta}\pm \overline{\beta}^2)^{-2/3} \alpha^{2/3} \beta^2$/2 \\
$b_F$ & $9N^{-2/3}/60$ & $d_5\, (d_4N)^{-2/3}$ & $2(\alpha^2 \beta^2 \overline{\beta}^4)^{-1/3} (5+4\beta^2 - 8\overline{\beta}^2)/30$ & $\frac{5\beta^2+4(3-2\beta^2\pm3\overline{\beta})}{30(\sqrt{2}\alpha)^{2/3} (\overline{\beta}\pm\overline{\beta}^2)^{4/3}}$ \\
\hline
\multicolumn{5}{c|}{Total energies}\\
\hline
$T$ & $N^2/4$ &  $-$ & $a^2 N[-N^2/3+N(\alpha_e - \alpha^2 (2\alpha_e)^{-1})-1/6]/2$ & $a^2 N(-N^2/3 + \alpha N/2 +1/12)/2$ \\ 
\hline
$T\TF$ &$T$ & $2d_4^{4/3}N^{7/3}/7$ & $a^2N(-N^2/3+\alpha N/2)/2 $ & $a^2N(-N^2/3+\alpha N/2)/2$\\
$T\one$ & 0 &  $-$ & $ a^2N[9N(8\alpha)^{-1}-1]/12$ & $a^2N/24$\\
\hline
$V$ & $T$ &  $-$ & $-\alpha^2 a^2 N[-N(2\alpha_e)^{-1} + 1]/2$ & $ -a^2 \alpha N(-N+2\alpha)/4$ \\
\hline
$V\TF$ & $T $ & $T\TF/2$ & $ -a^2 \alpha N(-N+2\alpha)/4$ &  $V$ \\ 
$V\one$ & 0 &  $-$ & $ -a^2N^2(32\alpha)^{-1}$ & 0 \\
\hline
$E$ & $2T$ &  $-$ & $-a^2 N(N^2/3 - \alpha_e N + \alpha^2 + 1/6)/2$&  $-a^2 N(N^2/3 - \alpha N + \alpha^2 - 1/12)/2$\\ \hline
$E\TF$ & $E$ & $3T\TF/2$ & $-a^2 N(N^2/3 - \alpha N + \alpha^2)/2$ & $-a^2 N(N^2/3 - \alpha N + \alpha^2)/2$ \\
$E\one$ & 0 &  $-$ & $a^2N[3N(4\alpha)^{-1}-1]/12$ & $a^2N/24$ \\
\hline
\end{tabular}
\end{center}
\caption{Potentials and useful formulas, both exact and semiclassical.  Notation:
$\tilde{j}=j-1/2$, $d_4 = 3 \pi^{3/2} \Gamma^{-2}(1/4) \approx 1.270820$, 
$d_5=1/5+d_4^{2}/27\approx 0.259814$,
$\alpha = \sqrt{2D/a^2}$, $\alpha_e = \sqrt{2D/a^2+1/4}$, $\beta = 1-\alpha N$, 
$\overline{\beta}=\sqrt{1-\beta^2}$. Dashes indicate no analytical result.}
\label{analy}
\end{table*}%

\begin{figure} [htb]
\centering
\includegraphics[width=\columnwidth]{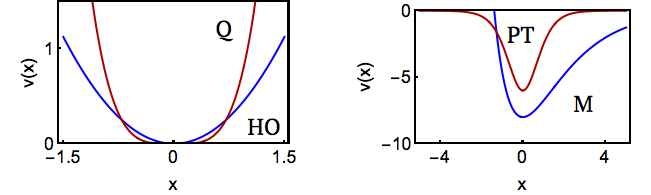}
\caption{Analytic potentials employed for the systematic investigation of
the uniform semiclassical approximations.}
\label{pot}
 \end{figure}

Several potentials are employed to illustrate our results.   Most have analytic forms
for at least their semiclassical quantities, and many have exact solutions.
All are infinitely differentiable.  These are illustrated in Fig. \ref{pot}.
For each form, we calculate both standard semiclassical quantities, and all the 
derived quantities used in this paper.  We also give the TF results.  These are
collected in Table \ref{analy}.

For any potential, the WKB energy components of each individual level can be immediately extracted if 
$\epsilon\sc_j[\beta v(x)]$ is known, where $\beta >0$.
Then
\ben
v_j = \frac{\partial \epsilon_j}{\partial \beta} \Big|_{\beta=1}
\label{vj}
\een
is the potential contribution, and $t_j$ may be found by subtraction. As indicated by
March and Plaskett\cite{MP56}, TF energies may be found by applying the 
Poisson summation formula to evaluate the corresponding sum of WKB energy components over occupied orbitals, and retaining only the average
contribution, e.g.,
\ben
T\TF = \int_{-\half}^{N-\half} dj\, t\sc_j
\label{TTFint}
\een
with similar forms for the other components. Because simple explicit results for WKB
energies are known for many potentials, the above allows a quick generation of TF energy components.

In those cases where analytic expressions are available for the exact energy components,
one can extract the $\gamma$-dependence of an energy component as:
\ben
T_\gamma (N,D,a) = T(N/\gamma,a/\gamma,D),
\label{Egam}
\een
where $D$ is the well-depth, and $1/a$ is a characteristic length scale (see Table \ref{analy}). By taking $\gamma\to 0$, the dominant term is given by TF,  and we denote the
next correction as $T\one$, e.g.,
\ben
T_\gamma \to T\TF/\gamma + \gamma\, T\one + ...
\label{Egamsmall}
\een
This is also listed Table \ref{analy}.  Table \ref{numt} gives numerical values for the constants associated to each potential we studied in this paper.

\sec{Leading corrections to Thomas-Fermi energies}
\label{lead}

\ssec{Pointwise accuracy}
\label{point}

Figures \ref{ng}, \ref{tg} and \ref{ngratio} demonstrate the much greater accuracy
of the semiclassical approximations relative to those of TF, {\em pointwise} in space.
For many applications, this is terribly important, such as e.g., in the calculation
of a surface energy.
However, in the world of DFT, there is overwhelming importance
given to knowing energies as functionals of the density.  So, while the semiclassical
formulas yield great improvements over the local approximation pointwise, what is
their performance for energies (which depend on global averages)?  Answering the above question is one of the main purposes of the present paper, and we show below the answer is more complicated than expected.
 
\begin{table}[htp]
\begin{center}
\begin{tabular}{|c|c|c|c|c|}
\hline
$\gamma$ & $\eta\sc$ & $\eta\TF$ & $\zeta\sc$ & $\zeta\TF$\\
\hline
1    & 0.0118 & 0.2552 & 0.1284 & 0.8445\\
1/2 & 0.0026 & 0.1558 & 0.0320 & 0.4164\\
1/4 & 0.0007 & 0.0913 & 0.0082 & 0.2063\\
\hline
\end{tabular}
\end{center}
\caption{Semiclassical and TF pointwise errors for the particle and kinetic energy densities of the SHO ($N=1$)}
\label{etaT}
\end{table}%

We begin to illustrate the issue by using the density- and kinetic energy-error 
measures introduced in Ref. \cite{RLCE15}, which are given by:
\bea
\eta&=&\frac{1}{N} \int_{-\infty}^{\infty} \mathrm{d}x |\tilde n(x) - n(x)|,\nonumber\\
\zeta&=&\frac{1}{T} \int_{-\infty}^{\infty} \mathrm{d}x |\tilde t(x) - t(x)|,
\label{eta}
\eea
where $\tilde n$ ($\tilde t$) indicates an approximate density (kinetic energy density), and $T$ is the exact kinetic energy.
The density-error measure is chosen so that it does not vanish too rapidly with increasing accuracy of the approximation, and is normalized to 2 if the approximate density has no overlap with the exact one.
Table \ref{etaT} shows these errors for a harmonic oscillator with unit frequency as a function of $\gamma$ for both
the particle- and the kinetic-energy densities as given by both TF and semiclassical theories. 
Even for $\gamma=1$, the semiclassical density error is
almost two orders of magnitude smaller than that of
TF, and vanishes much more rapidly with decreasing $\gamma$. 
The improvement in the kinetic energy density error is still
very good (the semiclassical error vanishes as $\gamma \rightarrow 0$, while TF does not), though not as spectacular at $\gamma=1$.
 
\begin{figure}[htp]
\includegraphics[width=\columnwidth]{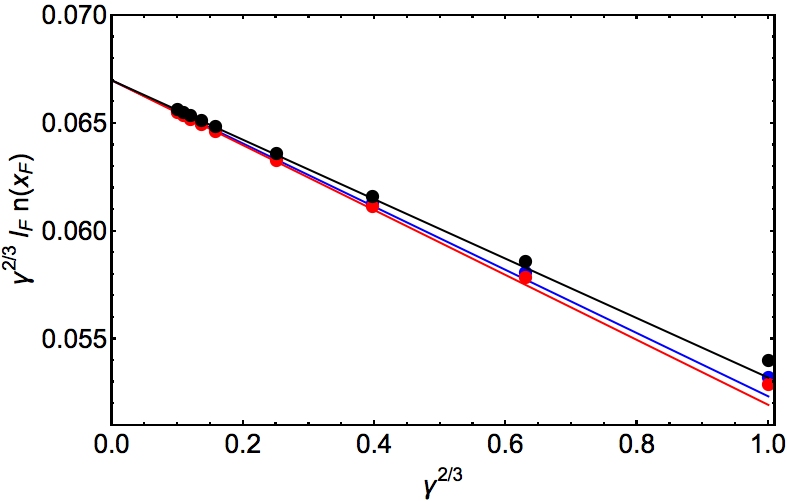}
\caption{Scaled dimensionless turning-point density,
$\gamma^{2/3} l\F n(x_F)$, as a function of $\gamma$ for
several potentials.  Points are exact, straight lines are
the leading behavior of the semiclassical formula,  Eq. \ref{ntsc}.
Black is harmonic oscillator, and blue and red are the left
and right turning points of a Morse potential of depth 6 ($a=1$).}
\label{tpd}
\end{figure}
Another beautiful illustration of the semiclassical formulas is seen by considering the
particle- and kinetic-energy densities at the turning point of the semiclassical
Fermi energy.  The explicit formulas are derived by expanding 
Eq. (\ref{nsc})
around a turning point.  This yields
\cite{KSb65, RLCE15, RB15}
\ben
l\F\n\sc\g(x\F)\to\frac{c_0}{\gamma^{2/3}} - c_1\, b\F,~~~
l\F^3 t\sc\g(x\F)\to -\frac{c_1}{6}, \label{sctp} 
\een
where 
\ben 
l\F=(2|v'\F|)^{-1/3}, 
\label{lF}
\een
$c_0=(12\pi^2 a_0^2)^{-1}$, $c_1=(2\pi\sqrt{3})^{-1}$,
and
\ben
b\F= \frac{1}{3}\left(\phi\F^2+ \frac{4\chi\F}{5} \right),
\label{bF}
\een
with
\ben
\phi\F=l\F^2\, \omega\F,~~~\chi\F= l\F^4 v''\F.
\label{phiFchiF}
\een
The length scale $l\F$ is that used by Kohn and Mattson\cite{KMd98} to
characterize the Airy gas.  It is the length scale defined by the
gradient of the potential at a turning point defined by the Fermi energy, and determines the
$\gamma\to 0$ behavior of both particle and kinetic densities at the turning point.
On the other hand, the leading correction to the turning-point
density contains two other scales: $\omega\F$, the classical frequency
of a trajectory at the Fermi energy, and $v''\F$, the second derivative of the potential at the turning point.  These can both be given in dimensionless quantities using $l\F$.  Thus $\chi\F$ is a dimensionless measure
of the second derivative of the potential and is a locally-determined
quantity (apart from the energy at which it is evaluated), while
$\phi\F$ is mixed, as it includes $\omega\F$, a global property, evaluated
on the local length scale, $l\F$.

{
\begin{table*}[htb]
\begin{center}
\begin{tabular}{| l | c | c | c | c | c | c | c | c |}
\hline
 & SHO & Q & PT & Morse & PT ($\delta$) & PT ($\delta$) & \multicolumn{2}{c|}{Double Well} \\
\hline
$v(x)$ & $x^2/2$ & $x^4$  & $D=21,a=1$ & $D=8, a=1/2$ & $\delta=1/8$ & $\delta=1/64$ & $b = 3.0$ & $b=3.3$\\ 
\hline
$\epsilon$ &  0.500  & 0.668 &  $-18.000$ & $-7.031$ & $-0.262$ & $-0.200$  & $1.29, 1.43$ & $1.52, 1.56$\\
$\epsilon^{sc}$ & 0.500 & 0.546 & $-17.884$ & $-7.031$ & $-0.191$ & $-0.133$ & 1.41 & 1.57 \\
\hline
$t$ & 0.250 & 0.445 & 1.385 & 0.469 & 0.107&  0.088 & 0.76, 1.21 & 0.77, 0.63\\
$t^{\rm WKB}$ & 0.250 & 0.364 & 1.495 & 0.469 & 0.154 &  0.129 & NC & NC \\ \hline
$v$  & 0.250 & 0.223 & $-19.384$ & $-7.500$ & $-0.370$ & $-0.288$ & 0.66, 0.84& 0.79, 0.88\\ 
$v^{\rm WKB}$ &0.250 & 0.182 & $-19.380$ & $-7.500$ & $-0.345$ & $-0.262$ & NC &NC \\
\hline
\multicolumn{9}{c|}{Fermi level}\\
\hline
$\epsilon_F $ & 1.000 & 1.376 & $-15.019$ & $-6.125$ & $-0.007$ & $-0.0001$ & 2.51 & 2.96\\
$\omega_F$ &  1.000 &  1.835 & 5.481  & 1.750 & 0.118 & 0.0155 & 1.45& 2.50 \\
\hline
$x_{F}$ & $-1.414$ & $-1.083$ & $-0.595$ & $-0.789, -1.323$ &$-2.939$ &  $-4.875$& $ - 2.120, - 0.087 $ & $ -2.270, -0.539$ \\
$l_F$ & $0.707$  &  0.462  & $0.315$ & $0.443,0.630$ & 3.305 & 12.765 & $0.375, 1.088$ &  $0.356, 0.576$ \\
$v_F''$ & $1.000$ &  14.079 & 4.373 & $11.684, 0.065$ & $-0.027$ & $-0.0005$ & $22.455,-4.455$ & $25.482, -3.702$\\
\hline
\multicolumn{9}{c|}{Dimensionless parameters}\\
\hline
$\phi_F$ & $0.500$  &  0.391& 0.543 & $ 0.343,0.695$ & 1.289 & 2.526 &  $0.204, 1.712$ & $0.318, 0.829$ \\
$\chi_F$ & $0.250$  & 0.639 & 0.043 &$0.450, 0.010$ & $-3.268$ & -12.762 & $0.443,-6.241$ & $0.411, -0.406 $  \\
$l_F^2\epsilon_F$ & 0.500 & 0.293 & $-1.488$ & $-1.202,-2.432$ & $-0.076$ & $-0.020$ & $0.353, 2.976$ & $0.376, 0.980 $\\
$b_F$ & 0.150 & 0.221& 0.110& $0.159,0.164$ &  $-0.442$ & $-1.275$& $0.186, -1.091$ & $0.216, 0.168$ \\
\hline
\multicolumn{9}{c|}{Regional energies}\\
\hline
$T^{\rm allow}$ & 0.293 & 0.519  & 1.618 & 0.547 & 0.113 & 0.089 & 1.37 & 1.65 \\
$T^{\rm allow, TF}$ & 0.250 & 0.393 & 1.453 & 0.458 & 0.113 & 0.087 & 1.15 & 1.41 \\
$T^{\rm allow, sc}$ & 0.266 &  0.401 & 1.542 & 0.484 & 0.113 & 0.084 & 1.22 & 1.50 \\
\hline
$T^{\rm forbid}$ & $-0.043$ & $-0.074$ &$-0.233$ &$-0.078$ & 0.000 & 0.000 & -0.33 & -0.24  \\
$T^{\rm forbid, TF}$ & 0.000 & 0.000 & 0.000 & 0.000 & 0.000 & 0.000 & 0.00 & 0.00 \\
$T^{\rm forbid, sc}$ & $-0.047$ & $-0.118$ & $-0.220$ & $-0.090$ & 0.000 & 0.000 & -0.01 & -0.25\\
\hline
$V^{\rm allow}$ & 0.185 & 0.123 & $-18.791$ & $-7.621$ & $-0.370$ & $-0.305$ & 1.27 & 1.29 \\
$V^{\rm allow, TF}$ & 0.250 & 0.197  & $-19.380$ & $-7.500$ & $-0.345$ & $-0.262$ & 1.56 & 1.67 \\
$V^{\rm allow, sc}$ & 0.182 & 0.122 & $-18.596$ & $-7.627$ & $-0.335$ & $0.249$ & 1.30 & 1.27 \\
\hline
$V^{\rm forbid}$ & 0.065 & 0.093 & $-0.593$ & $-8.009$ & 0.000 & 0.000 & 0.29 & 0.38 \\
$V^{\rm forbid, TF}$ & 0.000 & 0.000 & 0.000 & 0.000 & 0.000 & 0.000 & 0.00 & 0.00 \\
$V^{\rm forbid, sc}$ & 0.065 & 0.099 & $-0.622$ & $-8.009$ & 0.000 & 0.000 & 0.21 & 0.37 \\
\hline
\multicolumn{9}{c|}{Total energies}\\
\hline
$T$ & 0.250 & 0.445 & 1.385 & 0.469 & 0.107 & 0.088 & 1.21 & 1.41 \\ 
\hline
$T^{\rm TF}$ & 0.250 & 0.393 & 1.453 & 0.458 & 0.113 & 0.087 & 1.15 & 1.41  \\
$T^{\rm sc}$ & 0.218 & 0.283 & 1.322 & 0.419 & 0.112 & 0.084 & 1.04 & 1.25 \\
$T^{(1)}$ & 0.000 & NA & $-0.069$ & 0.010 & $0.0005$ & $-0.055$ & NC & NC\\
\hline
$V$ & $0.250$ & 0.223 & $ -19.384$ & $-7.500$ & $-0.370$ & $-0.288$ & 1.51& 1.67 \\
\hline
$V^{\rm TF}$ & 0.250 & 0.197& $-19.380$ & $-7.500$ & $-0.345$ & $-0.262$ & 1.56 & 1.67 \\ 
$V^{\rm sc}$ & 0.246 & 0.222 & $-19.219$ & $-7.508$ & $-0.335$ & $-0.249$ & 1.56 & 1.64\\
$V^{(1)}$ & 0.000 & NA &$-0.005$& 0.000 & $-0.028$ & $-0.031$ & NC & NC \\
\hline
$E$ & 0.500 & 0.668 & $-18.000$ & $-7.031$ & $-0.262$ & $-0.200$ &2.72 & 3.08\\
\hline
$E^{\rm TF}$ & 0.500 & 0.590 &  $-17.926$ & $-7.042$ & $-0.232$ & $-0.175$ & 2.71 & 3.08\\
$E^{\rm sc}$ & 0.464 & 0.505 & $-17.897$ & $-7.089$ & $-0.223$ & $-0.165$& 2.60 & 2.89\\
$E^{(1)}$ & 0.000 & NA & $-0.074$ & 0.010 & $-0.027$ & $-0.086$ & NC& NC\\
\hline
\end{tabular}
\end{center}
\caption{Numerical values of parameters for each potential used in this paper, 
with $N=1$ in each. When two values appear, they correspond to left- and right-turning
points respectively. NC denotes not calculated.}
\label{numt}
\end{table*}

In Figs \ref{tpd} and \ref{tpt}, we plot $n\sc\g(x\F)$ and $t\sc\g(x\F)$
for the harmonic oscillator and both sides of the Morse potential (with $N=1$).  A universal limit is clearly shown by Fig. \ref{tpd} which was first identified by Kohn and Sham\cite{KSb65}.
However, it also shows that the next correction obtained with the semiclassical uniform approximation is 
appropriate, and that these two terms alone yield results at $\gamma=1$
that are accurate to within 2\%.Table \ref{numt} indicates the Morse potential produces values of $b\F$ for the left and right turning points which are very close and only slightly larger than those of the HO.  Hence the bunching of the
straight lines in Fig. \ref{tpd}.

Our result, Eq. (\ref{sctp}), for
the kinetic-energy turning-point density, is likewise universal, but
does not produce the correct leading correction.
The derivation of  Ref. \cite{RB15} guarantees only that the leading terms
should be correct, and our figures are numerical illustrations of
this fact.  It is an added bonus that the leading correction is
also appropriate for the density.  That it is not so for the kinetic
energy density reflects the greater difficulty in achieving the
same number of terms in the expansion in $\gamma$ for the kinetic
energy, because of the two spatial derivatives in the operator.

\begin{figure} [htp]
\includegraphics[width=\columnwidth]{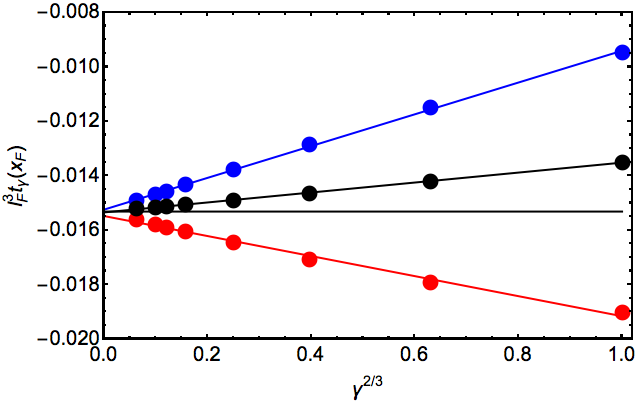}
\caption{Same as Fig. \ref{tpd}, but for the scaled dimensionless kinetic
energy turning-point density, $l\F^3 t\g(x\F)$.}
\label{tpt}
\end{figure}

\ssec{Regional particle numbers and energies}
\label{region}

\def\all{^{\rm allow}}
\def\for{^{\rm forbid}}
The previous subsection demonstrates the far greater accuracy of the uniform approximation relative
to TF, both at every point in space, and at a special one, the Fermi turning point.
But this is a measure-zero set, so that comparisons such as the
one just shown do not imply much about the accuracy of
expectation values over the entire system.  So next, as we do throughout this section, we separate errors in the traveling
region from those in the evanescent region.  We write 
\ben
f\all=\int_{x\F}^{x_{m}} dx\, f(x),~~
f\for=\int_{-\infty}^{x\F} dx\, f(x),
\label{regiondef}
\een
so that
\ben
\int_{-\infty}^\infty dx\, f(x) = f^{\rm forbid,L}+f^{\rm allow,L}+f^{\rm allow,R}+f^{\rm forbid,R}
\label{globaldef}
\een
where the superscripts denote the left and right contributions, respectively.
This is a basic method of capturing some fraction of real-space
information, while being able to connect to global integrals trivially.
Most importantly, it allows integration over the detailed quantum
oscillations introduced by the semiclassical approximation.

The behavior of the semiclassical expansion is very different in the
neighborhood of a turning point than either in the interior allowed
region or exterior forbidden. To analyze it, we note
that, as $\gamma\to 0^+$, the potential in the region where
$|\theta|/\gamma < \xi,$ with $\xi \rightarrow 0^+$.
(that is, in a small neighborhood around the the turning point) behaves almost linearly
and we can expand to second-order around the
Fermi energy turning-point:
\ben
v(x)=\epsilon_F+(x-x\F)v'\F + (x-x\F)^2 v''\F/2+..
\label{vexp}
\een
Defining
$ y = (x-x\F)/l\F$
it then follows that,
\bea
l\F\, k\F(y)& = &\sqrt{y}\left[1-\chi\F y/2 + O(y^2)\right],\nonumber\\
\theta\F(y) &=& \frac{2}{3} y^{3/2}\left[1- \frac{3\chi\F y}{10}+O(y^2) \right],\nonumber\\
z\F(y) &=  &y\left[ 1-\frac{2 \chi\F y}{10} + O(y^2) \right], \nonumber\\
\tau\F(y)/l\F^2  &= & 2y^{1/2}\left[1+\frac{\chi\F y}{6} + O(y^2)\right].
\label{vic}
\eea
Insertion of the above limiting forms in Eq. (\ref{nsc}) yields the following approximation
for the semiclassical densities
in the vicinity of a turning point:
\def\vic{^{\rm vic}}
\bea
\n\vic(x)& =& \frac{1}{l\F \pi} \left[\Kta(\theta\F)+ b\F\, \Kb(\theta\F)\right],\nonumber\\
t\vic(x)& =& \frac{1}{6\pi l\F^3} \left( \Kb(\theta\F) +y\left[\Kta(\theta\F)+ b\F\, \Kb(\theta\F)\right]\right).
\label{nvic}
\eea
There are many interesting aspects to these formulas.  First, they match the
the special case of a 1D Airy gas.
Second, we see it depends on the second derivative
of the potential at the turning point through the coefficient $b\F$. Further, taking the limit $y\to 0$
yields the previously given turning point results, Eq. (\ref{sctp}).

We next use these formulas to derive the leading corrections to TF
energies in specific spatial regions.  The key observation is that
the density difference from TF,
\ben
\Delta n(x) = n(x)-\n\TF(x)
\label{deltan}
\een
vanishes with some negative power of $\theta_F$ as $|x-x\F|$ 
becomes large in the allowed region. In the forbidden region
the decay is exponential. Thus integrals over moments of this quantity
are well-behaved and allows capture of the dominant energetic corrections
to TF. The same remains true for the kinetic energy density.

The particle number in a given region
itself provides a useful warm-up exercise.  Define
\ben
N\for = \int_{-\infty}^{x\F} \mathrm{d}x~ \n(x).
\label{Ndef}
\een
Inserting the turning point vicinity density, Eq. (\ref{nvic}),
and employing the simple relations $k\F\, l\F \approx {\sqrt{z\F}}$
(from Eq. (\ref{vic})) and
$k dx = \sqrt{z} dz$ (valid everywhere), the integrals can be performed analytically (see appendix),
to yield
\ben 
N\for\g \to N_0 - N_1\, \gamma^{2/3} + ..., 
\label{Ng}
\een
where 
\ben
N_0 = \frac{1}{6\pi\sqrt{3}} \sim 0.03063,~~~N_1=b\F\,\frac{a_0^2}{2}.
\label{N0}
\een
The first of these is a universal constant that applies to all turning
points, while the leading correction depends on specific details, through $b\F$.
Applying the same reasoning to the allowed region while assuming the contribution to the integrated functions at $x_m$ is zero (which is necessarily the case when $\gamma$ approaches zero), we find
\ben 
N\all\g \to \frac{N}{\gamma} - N_0 + N_1\, \gamma^{2/3}+....,
\label{dNfsc}
\een
i.e., term-by-term in the $\gamma$-expansion, the gain in particle number in the
evanescent region precisely cancels the loss from the allowed region.
This is an explicit demonstration
that the semiclassical density, which is {\em not} exactly normalized
in general, {\em is} normalized
to this order in the $\gamma$ expansion.

A less trivial piece of information is also contained in this result.  Since the TF density
is entirely within the allowed region, this shows that, to leading order, approximately
0.03063 electrons leak out beyond each turning point into the neighboring
evanescent region.  This is
a universal result for all 1D potentials, and is illustrated
in Fig. \ref{dNcf}. Even at $\gamma=1$, the integral of the density
in the forbidden region is close to the asymptotic
behavior contained in Eq. (\ref{Ng}).

\begin{figure}[htp]
\includegraphics[width=\columnwidth]{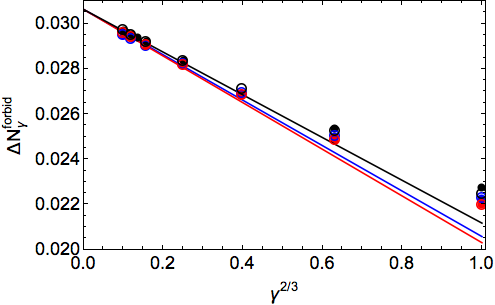}
\caption{$N^{\rm forbid}_{\gamma}$ for harmonic oscillator (black) and
left- (blue) and right- (red) turning points of
the Morse potential (all with $N=1$).  The straight lines are the
leading asymptotic behavior as $\gamma\to 0$, given by Eq. (\ref{Ng}). 
The empty circles correspond to results obtained with Eq. (\ref{ntsc}), while the filled are exact.}
\label{dNcf}
\end{figure}

The approach to the semiclassical limit is illustrated in Fig. \ref{dNcf}.
It corroborates the asymptotics given by Eq. \ref{Ng}. In particular, inclusion 
of the leading corrections to the semiclassical limit yields estimates for
the average number of particles in the forbidden region
which are accurate to within about 10\% for every case.

\begin{figure}[htp]
\includegraphics[width=\columnwidth]{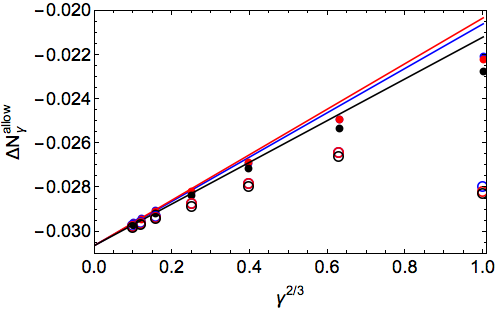}
\caption{Same as Fig. \ref{dNcf}, except the deviation from $N/\gamma$ in the
allowed region.}
\label{dNca}
\end{figure}
We also calculated the deviation from the TF result in the allowed region.  In the
asymptotic limit, this should become the mirror-image of the corresponding result in the
evanescent region.
This is shown in Fig. \ref{dNca}.  We find that while this is true for sufficiently
small $\gamma$, the deviations from straight-line behavior are much larger.  This
is because, e.g., for $\gamma=1$, the integral over the appropriate $K_j$ functions
are truncated at the mid-phase point, so that only a fraction of an oscillation
is included in the integration region.  On the other hand, as $\gamma\to 0$, the larger number of density oscillations
is averaged out, so the integral gets closer to its asymptotic value
as derived in the appendix.

\begin{figure}[htp]
\includegraphics[width=\columnwidth]{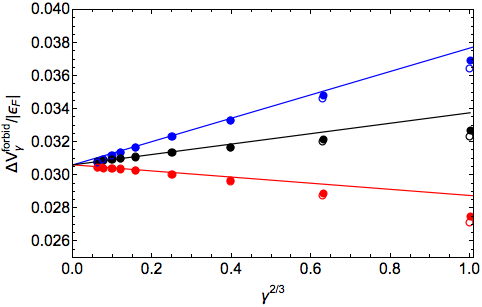}
\caption{Same as Fig. \ref{dNcf}, but for $V\for\g/|\epsilon_F|$.}
\label{dvcff}
\end{figure}

The next simplest observable we can address is the potential energy, which is directly
determined by the density.  A similar analysis yields:
\ben   
V\for\g \to -\epsilon\F N_0 +\gamma^{2/3} \left(\epsilon\F N_1 + \frac{a_0^2}{10l_F^2}\right) + ... 
\label{dvcf}
\een
Thus $V\for/\epsilon\F$ has a universal value ($-N_0$) for every potential.
The average potential
energy per electron in the evanescent region, is simply $-\epsilon\F$.
But the correction is system-dependent.  All this is illustrated in Fig. \ref{dvcff}.

\begin{figure}[htp]
\includegraphics[width=\columnwidth]{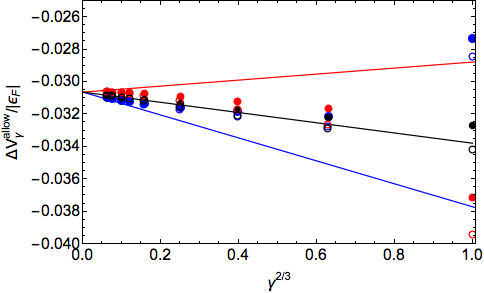}
 \caption{Same as Fig. \ref{dvcff}, but for the allowed region.}
\label{dvcaf}
\end{figure}
But, just as before, we find the change (relative to TF)
in the allowed region, shown in Fig. \ref{dvcaf},
exactly cancels the contribution from
the forbidden region, leaving zero contribution to the total potential energy, both for
the constant contribution and the $\gamma^{2/3}$ coefficient.  

\begin{figure}[htp]
\includegraphics[width=\columnwidth]{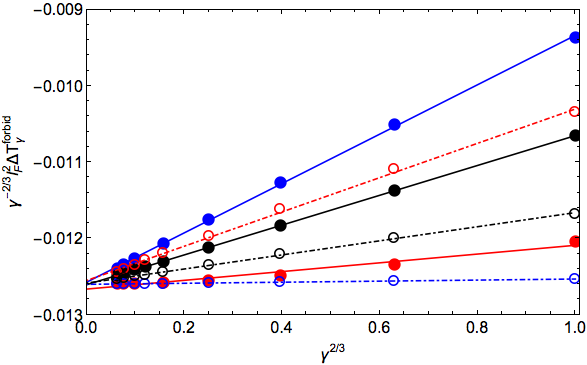}
\caption{Same as Fig. \ref{dNcf}, but for $l_F^2 \gamma^{-2/3} T^{\rm forbid}_{\gamma}$.}
\label{dtex}
\end{figure}
Finally, we repeat the calculation for the kinetic energy, finding
\ben
T\for\g \to -\frac{a_0^2}{10l_F^2}\gamma^{2/3}+... \label{dtcf}
\een
Note the important difference relative to the previous cases: there is no constant term in Eq. (\ref{dtcf}).
In Fig. \ref{dtex} we illustrate how this universal limit is approached
(in the evanescent region) for the Morse and harmonic oscillators. 
Here again, the error in the classically-allowed region is
cancelled by that of the classically-forbidden. 

\begin{figure}[htp]
\includegraphics[width=\columnwidth]{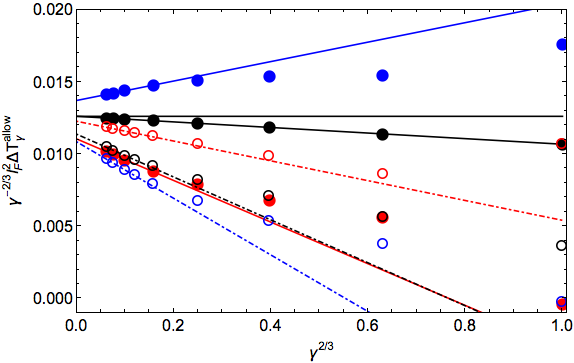}
\caption{Same as Fig. \ref{dtex}, but for the allowed region.}
\label{dTca}
\end{figure}

\ssec{Global energies}
\label{glob}

In this final subsection, we turn to global energy components.  
We have shown so far that, in each region of space, our semiclassical
formulas reproduce the leading corrections to TF observables, often in powers
of $\gamma^{1/3}$.  Now we add the corrections from each region to find
their effects on energy components.  As we have seen,
in the limit as $\gamma\to 0$, the corrections from each region cancel
each other, so no net contribution from the leading-order regional
contributions is left.

To understand this effect, we 
first consider the global integral of the density itself.
While TF is defined to be normalized by construction for any $v(x)$, the semiclassical
approximation is not.  The particle number $N$ enters only in defining $\epsilon\F$, but this does not guarantee normalization.  We can see this by adding Figs \ref{dNcf} and 
\ref{dNca} together.
While the leading corrections as $\gamma\to 0$ are given by the plotted straight lines, the deviation of the exact results from them is
much greater in the allowed region than in the forbidden, implying their sum is non-vanishing. In particular, the total
deviations are small and negative, 
as suggested by Fig. \ref{ngratio}, being never larger than 2\%. On the other hand, by construction, TF has
zero error for the particle number.

While all regional
contributions have $\gamma^{2/3}$ corrections to TF results, 
these have no effect in global properties, due to error cancellation.
We find that the leading corrections to TF energies scale linearly with $\gamma$
for each of the chosen potentials. The semiclassical approximations contain corrections of this order, but these are inaccurately given, as they have not been evaluated to the order needed.

\begin{figure}[htb]
\includegraphics[width=\columnwidth]{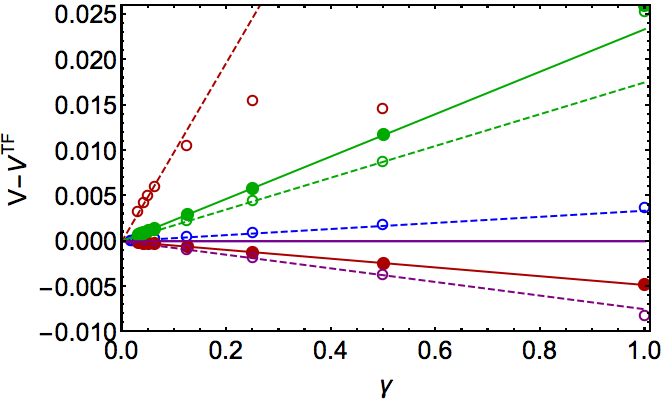}
\caption{$V-V^{TF}$, blue = SHO, green = quartic, red = PT, purple = Morse; empty = sc, filled = exact (when filled is not shown it is because it is zero).}
\label{Vtot}
\end{figure}
Fig. \ref{Vtot} shows the deviation from TF values for the wells given in Table \ref{analy}, for both semiclassical and exact calculations.
It also includes the leading corrections, which are given as straight
lines.  For the semiclassical approximation, these are numerical fits to a subset of the plotted points.  In the case of the exact calculations, the slopes provide leading corrections
to TF as determined from the exact analytic energy components
given in Tables \ref{analy} and \ref{numt}.
We see that in some cases, the
corrections to TF are positive, while in others negative corrections are found.  On the other hand, we see that the semiclassical approximation
produces linear corrections, but these are not accurate.  Moreover, the
errors are not particularly systematic. Inclusion of higher-order
corrections in terms of the potential might improve agreement with the
exact curves, but that is beyond the scope of the present study.

\begin{figure}[htb]
\includegraphics[width=\columnwidth]{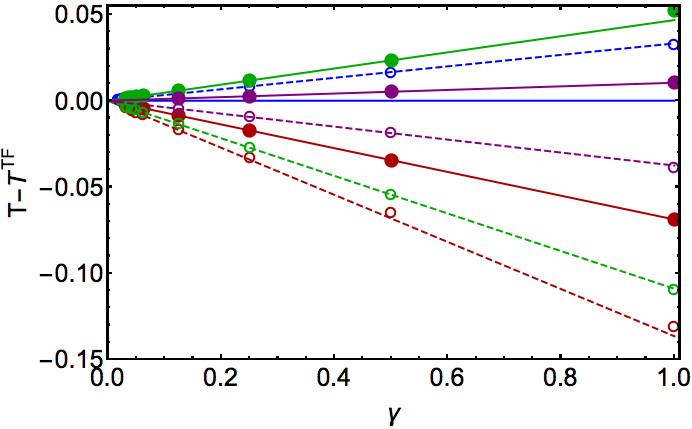}
\caption{$T-T^{TF}$, blue = SHO, green = quartic, red = PT, purple = Morse; empty = sc, filled = exact (when filled is not show it is because it is zero).}
\label{Ttot}
\end{figure}
In Fig. \ref{Ttot}, we find similarly baleful results for the deviations
of the kinetic energy from the TF predictions.  Note that these errors are
typically about 5 times larger than those for the potential energy.
Thus the errors for the total energy look very similar to this figure.
We also note that, if we had an approximation that yielded the slopes of these curves,
it would generically be very accurate, even for $\gamma=1$.  Since the TF errors
change sign with different potentials, no simple gradient expansion or even a generalized
gradient approximation, can hope to yield accurate corrections to TF in this limit.
Such corrections remain tantalizingly out of reach at present.

We conclude that although the semiclassical uniform approximation improves
over TF for all points in space, and even for each region individually as
defined earlier, it does {\em not} yield the leading corrections to the
total energies of the system.

\sec{Beyond generic situations}
\label{beyond}

The potentials we have studied so far have been generic situations of
particles
in infinitely-differentiable wells.  
We have included several different situations
(symmetric versus asymmetric, harmonic versus quartic)
to illustrate and analyze
the general behavior of the semiclassical uniform approximation especially trying
to bridge the gap between
its pointwise accuracy but its poorer
energetic performance.  The current section
probes more unusual cases (one can think of many) to see what happens.

\ssec{Limitations of semiclassical approximations}
\label{breaksc}

In this subsection, we explore various breakdowns of the semiclassical
approximations. Before doing so, we stress that, for sufficiently smooth
potentials, the semiclassical results {\em always} become relatively
exact everywhere as $\gamma\to 0$.  Here, however, we explore its
application at $\gamma=1$, i.e., not as an approach to the semiclassical
limit, but instead finding systems where either the semiclassical approximations cannot be applied at all
because the classical motion is unbound, or it can be used but yields
relatively inaccurate results.  The latter can happen
if the rate of change of the potential at a Fermi energy turning point is 
different from zero, but very small. 

Just because the semiclassical formulas can be considered as expansions
around the TF limit, there is no reason to assume they are well-behaved for all potentials.
We have already mentioned the fact that in general it is not
normalized.  We also note that it is not even guaranteed to yield
densities that are positive everywhere, as the oscillations in $\Kb$
yield negative contributions (which are much smaller than the
positive contributions of $\Kc$ everywhere unless $l\F \rightarrow \infty$). 

\sssec{Smooth potentials that vary too rapidly}
\label{smooth}

A relatively trivial case that illustrates the generic breakdown of
both the TF and semiclassical approaches is found by squeezing 
a continuous and well-behaved potential well so that it approaches
a delta-well.  The simplest example is the PT well, in which we take
the limit where $a \to \infty$, with $D=Z/(2a)$.  With this choice, the PT
potential approaches $v(x)=-Z\delta(x)$.  But we require 
\ben
\beta = 1 - Na/{\sqrt{2D}}
\label{betaD}
\een
to remain positive in order to define a semiclassical Fermi energy, and
thus a TF or a semiclassical solution (Table \ref{analy}).  For $a > 1$, the potential
no longer bounds even one semiclassical solution, and both TF and
our semiclassical approximation do not exist.

\sssec{Hard wall limit}
\label{hard}

Another limit of considerable interest is that of hard walls, which was
studied extensively in earlier work \cite{CLEB10}. A simple method
to obtain potentials of this class is by considering a symmetric potential
which on the left satsifies
\bea
v(x)&=&-F(x+L/2),~~~ x < -L/2\nonumber\\
&=& v^{box}(x),~~~ x > - L/2,
\eea
where $F$ is a constant which determines the slope of the potential,
and $v^{box}(x)$ vanishes at $-L/2$ and is smooth.  If
we take the limit where $F\to\infty$ and $v^{\text{box}}=0$, we recover a flat box.
The Fermi energy is well-behaved in this example.  However, 
$\n\sc(x\F)$ (and its kinetic counterpart) is not.  As $F\to\infty$,
$l\F\to 0$, so that $\n\sc\F(-L/2) \to \infty$, which is highly unphysical.
The reason this happens is because of the kink in the potential
at $-L/2$.  As $F$ grows, the exact density becomes {\em smaller}
at the turning point, while the semiclassical formula diverges instead. 

\sssec{Top of the well}
\label{top}

Two basic assumptions of the semiclassical approximations are that 
neither $\omega\F$, nor $v'\F$ vanish.
Thus we expect the approximations to perform badly whenever
these quantities become too small.  This happens, e.g., if the semiclassical
Fermi energy is close to zero for a potential that vanishes
at large distances.

\begin{figure}[htb]
\includegraphics[width=\columnwidth]{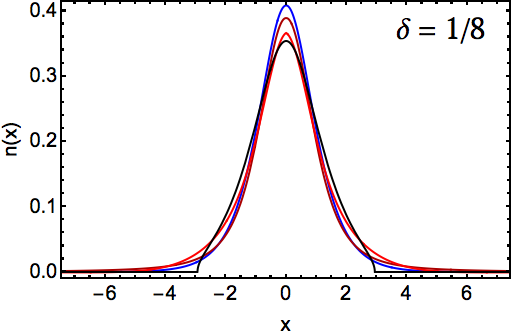}
\caption{Densities for PT with $D=1/2+1/8$. Blue is exact,
black is TF, red is semiclassical, while dark red is semiclassical
with just the first contribution.}
\label{del8}
\end{figure}

\begin{figure}[htb]
\includegraphics[width=\columnwidth]{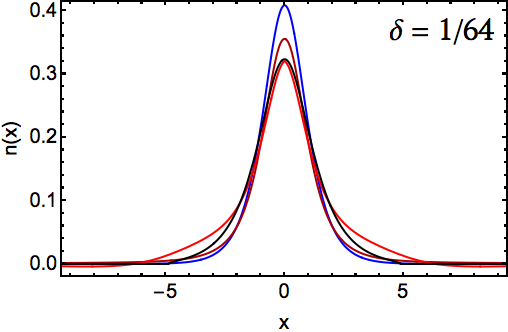}
\caption{Same as Fig. \ref{del8}, but with $D=1/2+1/64$.}
\label{del64}
\end{figure}

To create such an instance, we note that, while the PT potential
always binds a particle when treated exactly, the semiclassical Fermi energy vanishes as $D\to 0$ (see Table \ref{analy}).  Thus, we write $D=1/2 + \delta$,
and consider what happens as $\delta\to 0$.  Figs. \ref{del8} and \ref{del64}
tell the story. Even at $\delta=1/8$, one already sees a slightly negative
contribution to the density out in the tail, due to the second term in 
Eq \ref{nsc}.   We can also see that for most values of $x$, the semiclassical
approximation is worse than TF.  On the other hand, if we neglect $K_1$ in Eq. (20), we find a much more accurate density, without the local over and underestimates of the complete semiclassical approximation.  Thus in this
limit, it is better to ignore the $K_1$ corrections, but we note that the $K_1$ corrections
must be included to create a uniform approximation. For sufficiently
small $\gamma$, even for $\delta=1/64$, the uniform approximation will become exact pointwise.

\ssec{Extended systems}
\label{extend}

There is of course tremendous interest in applications of DFT to extended systems.
In this section, we derive the limit of our result that applies to such systems.

For any potential of the type we have considered so far, we create a new one
defined as:
\bea
v\L(x)&=&v(x),~~~x < x_m\nonumber\\
&=& v(x_m),~~~x_m < x < x_m+L\nonumber\\
&=& v(x-L),~~~ x > x_m+L.
\label{vext}
\eea
Note that this generalization depends on the value of $x_m$ which is a function
of the Fermi energy.  In fact, for our purposes, we start with a given particle
number $N$ in the original well, which defines $\epsilon\F$.  We then chose $M > N$
as the number of particles in $v\L(x)$, and this defines
\ben
L = (M-N) \pi/k\F(x_m).
\label{Lext}
\een
By this choice, the Fermi energy of the system containing $M$ particles in $v\L(x)$ matches that of
the $N$ particles in $v(x)$.  We now consider what happens as $M\to\infty$, so $L$
does also.  In the vicinity of the left turning point, $\alpha\F$ is always very small,
as $\tau\F(x) << T\F$.  Thus $\sin\alpha\F \to \alpha\F = \omega_F \tau_F(x)$. It follows that
\ben
d\F(x) \to \tau\F(x)\, k\F(x)
\label{surf}
\een
This is the only change needed to apply Eq. (\ref{nsc}) to a surface problem, with a fixed chemical
potential.  The distant turning point has become irrelevant.

It is beyond the scope of this work to explore these surface situations, but 
it is of considerable interest to compare the results with e.g., those of Mattson
and Kohn, and the subsequent development of density functionals using the Airy
gas \cite{LMA14}.  Our formulas yield quantum oscillations that typically extend deeply
into the bulk (i.e., for $x > x_m$) and moreover, as $\gamma\to 0$, the
results in a neighborhood of the surface depend on $v\F''$, i.e., the Fermi-level
curvature.  On the other hand, as we have seen, none of these effects necessarily
contribute to the energy.

\ssec{Tunneling}
\label{tunnel}

Here, we generalize our approximations to the case where more more than two turning points exist for the Fermi energy, by simply
treating each well independently.  Our aim is to study the accuracy
of the semiclassical formulae for densities penetrating barriers.  A simple case is
given by a double well with
\ben
v(x)=\frac{1}{2}\left(x^2-\frac{b^2}{4}\right)^2,
\label{vdw}
\een
with $b$ positive.  
In particular we check the local and energetic corrections to
TF for two ``bond lengths"(distance between the
minima of the studied potential).

\begin{figure} [htb]
\includegraphics[width=\columnwidth]{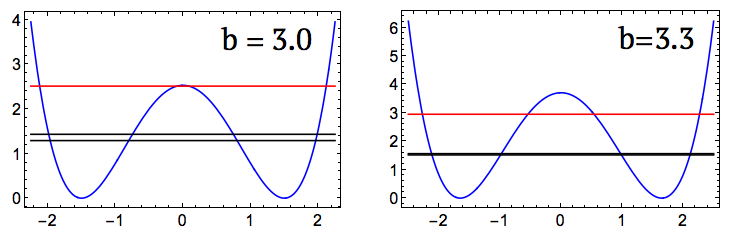}
\caption{Double well potentials (blue) with near degenerate eigenstates (black)
and Fermi level (red) for $b=3.0$ in Eq. (\ref{vdw}) and for $b=3.3$.}
\label{dwB}
\end{figure}
\begin{figure} [htb]
\includegraphics[width=\columnwidth]{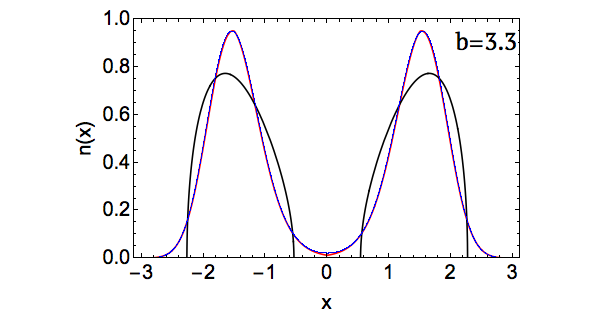}
\caption{Particle density (blue is exact, red is semiclassical, black is TF)
for the double well of Fig. \ref{dwB} with $b=3.3$.}
\label{ndwB}
\end{figure}
\begin{figure} [htp]
\includegraphics[width=\columnwidth]{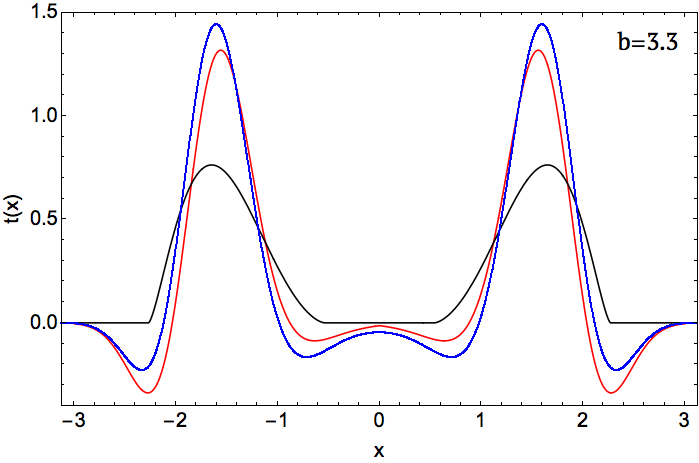}
\caption{Same as Fig. \ref{ndwB}, but for kinetic energy density.}
\label{tdwB}
\end{figure}

In Fig. \ref{dwB}, we show a generic case of a potential with a substantial
barrier separating two wells. On the r.h.s $b=3.3$, and the exact eigenvalues are almost degenerate,
while the semiclassical Fermi energy is not close to the top of the barrier.
Fig. \ref{ndwB} shows our usual excellent results for the densities
in this case.  In the evanescent region near each turning point, the
semiclassical formula is extremely accurate.  It fails only near the
midpoint, $x=0$, as expected.  At $x=0$ and
nearby, contributions from left and right wells interfere with each other, and our
approximation does not include such effects.  In Fig. \ref{tdwB}, we see that
the semiclassical kinetic energy density shares these traits, though deviations from the exact result are larger, as
expected.

\begin{figure} [htp]
\includegraphics[width=\columnwidth]{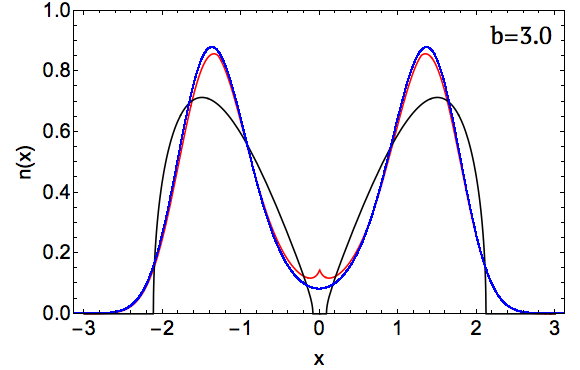}
\caption{Same as Fig. \ref{ndwB}, but for $b=3.0$.}
\label{ndwA}
\end{figure}

\begin{figure} [htp]
\includegraphics[width=\columnwidth]{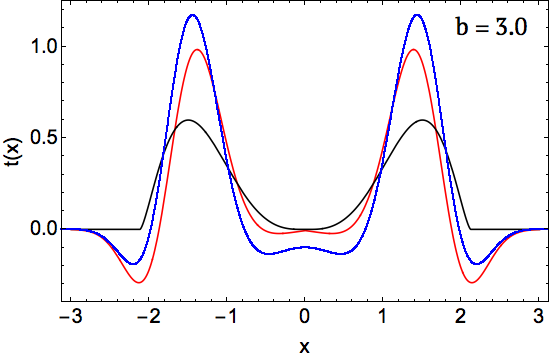}
\caption{Same as Fig. \ref{tdwB}, but for $b=3.0$.}
\label{tdwA}
\end{figure}

Even when we reduce the separation $b$ to a smaller distance, pushing $\epsilon\F$
to almost the top of the barrier, as shown on the l.h.s of Fig. \ref{dwB},
thus creating the conditions of Sec \ref{top}, the semiclassical
approximations continue to perform well.  The density is plotted in Fig. \ref{ndwA},
and comparable quality is achieved for the kinetic energy as is shown in Fig. \ref{tdwA}.

\sec{Conclusions}
\label{conc}

This paper explores in considerable detail the performance of the uniform
semiclassical approximations for the particle- and kinetic-energy densities
first presented in Ref. \cite{RLCE15}. A large variety of properties have been found analytically
and been tested numerically for a diverse set of potentials (all infinitely differentiable).
By defining regional contributions to particle number and energy components, 
we show that the spectacular pointwise accuracy of the semiclassical approximations
does transfer to integrals over spatial domains. But we also showed how the 
contributions from the semiclassical corrections vanish when integrated globally,
so that these contributions are {\em not} the leading corrections to energy components.

There are many extensions of this work that have only been mentioned or
superficially pursued here.  
One such is the derivation of semiclassical approximations which 
accurately describe the tunneling region between different wells. This should be uniformly accurate
for all values of $x$. Another concerns a more detailed study of double wells, e.g., checking how accurate
the uniform semiclassical approximations are in the multiple well case relative to the case of a single. 
An additional relevant extension of this work could be obtained by generalizing the treatment here given to
3D systems that are uniform in two directions, but vary in a third.
Such would be useful for the study of quantum dots, cold-atom fermion traps, and
plasma physics (when generalized to high temperatures) \cite{CJ15}.
Yet one more important area is surface energies.  We have generalized but not tested
the corresponding semiclassical approximations for such problems, where regional properties do matter very much.

Lastly, we discuss the relevance of these results to density functional
theory\cite{DG90}. Of greatest interest is the pursuit of a non-interacting
kinetic energy functional of the density, often
referred to as allowing orbital-free DFT \cite{WC00}.  In general, it has been found
that, for interacting 3D electronic systems, the 2nd-order gradient expansion
for the kinetic energy is highly accurate \cite{DG90}, but higher orders often worsen
properties \cite{LCPB09}.  Here, we have shown that functionals of the potential that go
beyond the local approximations can be extremely accurate pointwise,
while showing no systematic improvement globally.  This indeed mimics
experience in the real 3D world, where pointwise agreement can be greatly
improved in 4th order, without corresponding global improvement \cite{PC07}.
All this deepens further the mystery of the difficult relationship between potentials
and densities for fermionic systems \cite{L81}.
We have also found that, in order to recover the leading regional corrections
to TF results, the slope of the potential at the turning point is insufficient,
and that the next derivative contributes substantially to these corrections.  This appears important
for the construction of approximate functionals using the Airy gas \cite{KMd98, LMA14, AM05}.
Again, there are obvious further extensions of this work.  For example,
does the local density approximation, applied to either the exact or 
the semiclassical density, yield more accurate pointwise and regional properties
than when applied to its self-consistent density?  Can any of the potential functional
approximations given here be converted to simple density functional forms, that capture
their energetic consequences?

\begin{figure}[htb]
\includegraphics[width=\columnwidth]{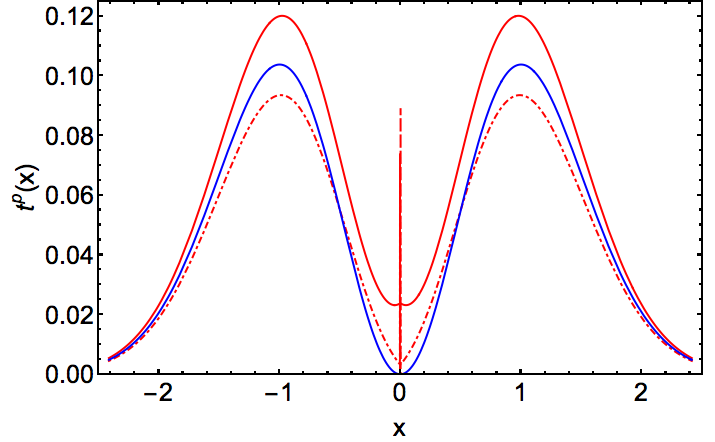}
\caption{Positive kinetic energy $t^p(x)$: Exact (blue), 
semiclassical (dot-dashed red), semiclassical without higher-order
terms in $\gamma$ (continuous red).}
\label{tposF}
\end{figure}
Finally, we mention the choice of kinetic energy density used in this work.
Our choice is a natural one starting from semiclassical analysis
since it assigns a local orbital kinetic energy $|\phi_i(x)|^2 p_i^2(x)/2$, where $p_i(x) = \epsilon_i-v(x)$, to each occupied state of the system. Thus it is directly analogous to the classical kinetic energy density of a configuration space distribution of particles with energy $\epsilon_i$.

But the purely positive choice has become popular in DFT, especially
for the construction of meta-GGA's \cite{SRP15}. They are related via
\ben
t^{\rm pos}(x) = \half \sum_{i=1}^N \int dx \left|\frac{d\phi_i}{dx}\right|^2 = t(x) + 
\frac{1}{4} \frac{d^2\n}{dx^2}.
\label{tpos}
\een
Thus we can use Eq. (\ref{nsc}) to construct approximations for
$t^{\rm pos}(x)$, and these are plotted for the harmonic oscillator with 1 particle
in Fig. \ref{tposF}. We see that the accuracy of the approximations to
$t^{\rm pos}(x)$ is similar to those for $t(x)$ in Fig \ref{tg}, apart from the
mid-phase point. There the small kink in the semiclassical approximation for $n(x)$ has its effects magnified by the two derivatives
of the density which must be taken in order to evaluate $t^{\rm pos}(x)$.
However, the resulting analytic expression for the positive kinetic energy density
includes derivatives up to second-order in the potential, unlike
the expressions of Eq. (\ref{nsc}) which contain none.
These can be shown to be irrelevant as $\gamma\to 0$. Furthermore, because $n^{sc}$ and $t^{sc}$ are only guaranteed to include the leading corrections to TF, the terms involving high powers of $\gamma$ which emerge from $\mathrm{d}^2n^{sc}(x)/\mathrm{d}x^2$ are likely incorrect. Thus, we also construct an approximation for the positive kinetic energy density in which
all such higher-derivative terms are neglected.  This is shown by the dashed
line in Fig. \ref{tposF}. As expected, it is substantially more accurate than the semiclassical approximation obtained for $t^{\rm pos}$ which does not remove higher-order terms of the $\gamma$ expansion. Hence use of $t^{pos}(x)$ in this way either involves
more complicated analytic expressions or a loss of accuracy.  

To summarize, we have explored the accuracy of uniform semiclassical approximations
to both the particle and kinetic energy densities of noninteracting fermions in 1d, finding (unfortunately) that
their spectacular performance pointwise in improvement over TF results does 
{\em not} translate directly to improved energy components.  This is explained
by the difference between regional corrections to the energy and global
averages.  Limitations of the semiclassical approximations have been explored and the relevance to
density functional theory discussed.

We gratefully acknowledge support of the NSF through grant number CHE-1464795.

\sec{Appendix - Relevant Properties of Airy functions}
\label{append}
\ssec{Definitions}
\label{defs}
We define the following notation for pairs of Airy functions
\ben A_0 = - a a', ~A_1 = a^2, ~A_2 = a'^2, \een
where
$a(z) = Ai(-z)$ and $a'=da/dz$.  From these, we define
the combinations relevant to this work:
\bea
\Kta(z)&=&\pi \left[ z A_1(z) + A_2 (z)\right],\nonumber\\
\Kb(z)&=&\pi A_0(z),\nonumber\\
\Kc(z)&=&z^{-3/2}\left[z\Kta(z)-\frac{\Kb(z)}{2}\right],
\eea

Next we list the asymptotic expansions of $A_i(z)$ in the
three regions with qualitatively different behavior.
In the traveling region,
\bea
\pi z^{\pm 1/2} A_{1,2}(z) &\to&
\frac{1\pm \text{sin}(2\theta)}{2}-\frac{(1\mp 6 )\text{cos}(2\theta)}{72\theta}
+...,\nonumber\\
\pi A_0(z) &\to& - \frac{\text{cos}(2\theta)}{2} +
\frac{6+ \text{sin}(2\theta)}{72\theta} + ...
\label{A_i_trav}
\eea
In the classically-forbidden region
\bea
|z|^{\pm 1/2} A_{1,2}(z)&\to& \frac{e^{-2|\theta|}}{4}
\left(1+ \frac{1\mp 6}{36|\theta|}+
O(|\theta|^{-2}) \right), \nonumber\\
\pi A_0(z) &\to& -\frac{e^{-2|\theta|}}{4}\left(1 + \frac{1}{36|\theta|}
+O(|\theta|^{-2}) \right).
\label{A_i_evan}
\eea
In a small neighborhood of some turning point,
\bea
A_1(z) &\to& a_0^2 + \frac{z}{\pi\sqrt{3}}+O(z^2),\nonumber\\
A_2(z) &\to& \frac{1}{12\pi^2 a_0^2} -
\frac{z^2}{2\pi \sqrt{3}} +O(z^3),\nonumber\\
A_0(z)&\to& -\frac{1}{2\sqrt{3}\pi} 
+ \frac{z}{12\pi^2 a_0^2} 
+O(z^2),
\label{A_i_tp}
\eea
where
\ben a_0 = a(0) = \frac{3^{-2/3}}{ \Gamma(2/3) }=0.35028.  \een

Now we combine the results above to obtain asymptotic approximations to the $K_i$ functions 
In the traveling region,
\bea 
\Kb(z)&\to& -\frac{\text{cos}(2\theta)}{2}+ \frac{6+\text{sin}(2\theta)}{72\theta} + O(\theta^{-2}),\nonumber\\
\Kc(z)&\to& 1 + O(\theta^{-2}),\nonumber\\
\Kta(z)&\to& \left(\frac{3\theta}{2}\right)^{1/3}\left[1 - \frac{\text{cos}(2\theta)}{6\theta} + O(\theta^{-2})\right]. 
\label{K_i_trav}
\eea
In the evanescent,
\bea
\Kb(z) &\to&-\frac{e^{-2|\theta|}}{4}
\left[1+\frac{1}{36|\theta|} + O(|\theta|^{-2})\right], \nonumber\\
\Kc(z) &\to& -\frac{e^{-2|\theta|}}{2}\left[1+O(|\theta|^{-2})\right],\nonumber\\
\Kta(z) &\to& \frac{e^{-2|\theta|}}{2} \left(\frac{3|\theta|}{2}\right)^{1/3}\left[1-\frac{1}{6|\theta|} + O(|\theta|^{-2})\right] . 
\label{K_i_evan}
\eea
Near a turning point,
\bea
\Kb(z)&\to& -\frac{1}{2\sqrt{3}}+\frac{z}{12 \pi a_0^2}+O(z^2),\nonumber\\
\Kc(z)&\to& z^{-3/2}\left[\frac{1}{4\sqrt{3}}
+\frac{z}{24\pi a_0^2} +O(z^2) \right],\nonumber\\
\Kta(z) &\to&\frac{1}{12\pi a_0^2} +\pi a_0^2 z +O(z^2).
\label{K_i_tp}
\eea
These were used to derive Eq. (27) of the text.

Finally, in obtaining corrections to TF in the 
forbidden regions,
we evaluated a variety of integrals of Airy functions. Here we
show those which are needed to verify our results. Defining
\ben
I_p^m = \int_0^{\infty} \mathrm{d}z ~z^{m}K_p(z),
\label{Ipmdef}
\een
then it follows that \cite{VS04},
\ben
I_2^0=-\frac{1}{6\sqrt{3}}, ~I_1^0 = \frac{\pi a_0^2}{2},
~I_2^1=\frac{I_1^0}{5}, ~I_1^1 = -\frac{1}{24\pi a_0^2}.
\label{Ipmres}
\een
To repeat the calculation in the allowed region, one must subtract
the corresponding TF values (the dominant terms in Eq. (\ref{K_i_trav})) from the
$K_i$ functions to find the deviations from TF.
These allow verification of Eqs. (\ref{Ng}-\ref{dtcf}) of the text.

\bibliography{MasterOld}

\label{page:end}
\end{document}